# The ECLAIRs telescope onboard the SVOM mission for gamma-ray burst studies

Stéphane Schanne *on behalf of the ECLAIRs collaboration*[1]

*CEA Saclay, DSM / IRFU (former DAPNIA) / SAp, 91191 Gif sur Yvette, France*

**Abstract.** The X- and gamma-ray telescope ECLAIRs onboard the future mission for gamma-ray burst studies SVOM (Space-based multi-band astronomical Variable Objects Monitor) is foreseen to operate in orbit from 2013 on. ECLAIRs will provide fast and accurate GRB triggers to other onboard telescopes, as well as to the whole GRB community, in particular ground-based follow-up telescopes. With its very low energy threshold ECLAIRs is particularly well suited for the detection of highly redshifted GRB. The ECLAIRs X- and gamma-ray imaging camera (CXG), used for GRB detection and localization, is combined with a soft X-ray telescope (SXT) for afterglow observations and position refinement. The CXG is a 2D-coded mask imager with a 1024 cm$^2$ detection plane made of 80×80 CdTe pixels, sensitive from 4 to 300 keV, with imaging capabilities up to about 120 keV and a localization accuracy better than 10 arcmin. The CXG permanently observes a 2 sr-wide field of the sky and provides photon data to the onboard science and triggering unit (UTS) which detects GRB by count-rate increases or by the appearance of a new source in cyclic sky images. The SXT is a mirror focusing X-ray telescope operating from 0.3 to 2 keV with a sensitivity of 1 mCrab for 100 s observations. The spacecraft slews within $\simeq$3 min in order to place the GRB candidate into the 23×23 arcmin$^2$ field of view of the SXT, after which it refines the GRB position to about 10 arcsec. GRB alerts are transmitted to ground-observers within tens of seconds via a VHF network and all detected photons are available hours later for detailed analysis. In this paper we present the ECLAIRs concepts, with emphasis on the expected performances.

## INTRODUCTION

Gamma-ray bursts (GRB) are cosmic events, which are seen in space borne gamma-ray detectors as count rate increases during short periods of time, from tens of milliseconds to minutes. They are the signature of very energetic explosions in the Universe, mostly occurring at cosmological distances, and believed to be linked to the formation of black holes. The most popular models assume the collapse of a rotating massive star (for long duration GRB) or the merger of two neutron stars (for short GRB). Following the Gamma-ray event, afterglows are often detectable in other wavebands (visible, radio, X-rays). Those provide crucial additional information to study the physics of the event itself (constrain the GRB models, study their relativistic jets and shocks, study the link between GRB and supernovae), to determine the distance of the event (measuring the redshift of its host galaxy), to use the event as a background light source to study the foreground universe, or to use collections of those events as tools for cosmology (study the star formation history, the very first stars, constrain the cosmological parameters), or to solve questions of fundamental physics (GRB as sources of ultra-high energy cosmic rays or gravitational waves).

---

[1] CEA Saclay (Commissariat à l'Energie Atomique), CESR Toulouse (Centre d'Etude Spatiale des Rayonnements), APC Paris (Astro-Particules et Cosmologie), IAP Paris (Institut d'Astrophysique de Paris), IASF Milano (Istituto di Astrofisica Spaziale e Fisica Cosmica), MIT Boston (Massachusetts Institute of Technology), CNES Toulouse (Centre National d'Etudes Spatiales)

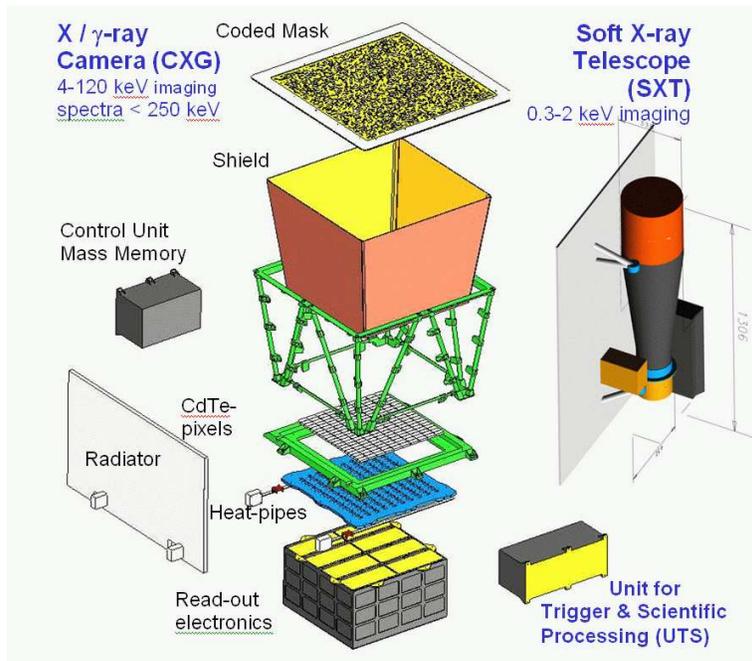

**FIGURE 1.** Components of the ECLAIRs telescopes CXG and SXT

Currently the satellites INTEGRAL and Swift deliver most of the GRB triggers to the ground-based observers, among which robotic follow-up telescopes, which refine the space-given localization to a precision matching the small fields of view of the large spectroscopic 8-m class telescopes. In 2013 the ECLAIRs gamma-ray detector, foreseen to fly on a low-earth-orbit satellite, is expected to carry on the hunt and deliver $\simeq 80$ GRB triggers per year to the world-wide community of observers.

## CONCEPT OF ECLAIRS

The ECLAIRs project, managed by CEA Saclay, France, is currently in its detailed technical design phase, with the development phase expected to start in 2008. Unlike the previous project versions [1, 2], ECLAIRs will now be part of the payload of the SVOM mission (Space based multi-band astronomical Variable Objects Monitor) [3], an approved scientific satellite, which is developed in collaboration between French, Italian and Chinese partners, with a launch date scheduled in 2013.

The ECLAIRs flight hardware (Figure 1) is composed of a 2D-coded mask aperture telescope CXG (camera for X- and gamma-rays), similar to the imager ISGRI onboard the INTEGRAL satellite. The detection plane (36 cm side length) is composed of $80 \times 80$

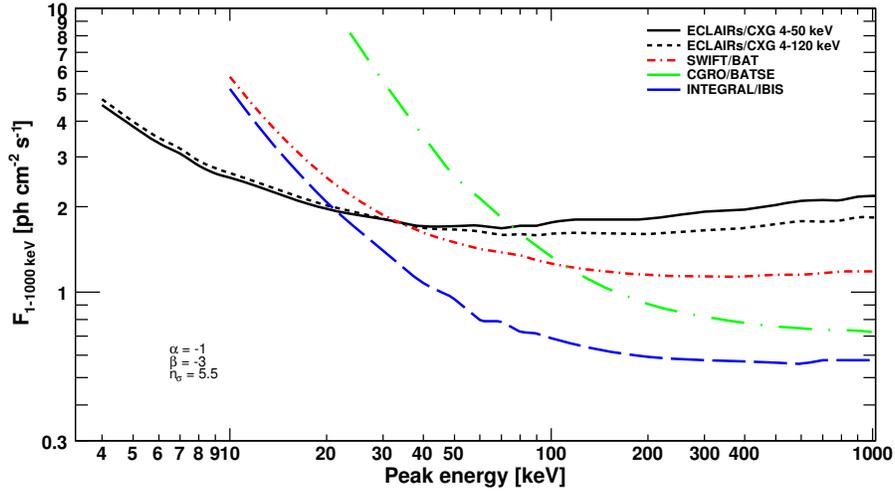

**FIGURE 2.** Source flux reconstructed by the localization algorithm (for a source at detection limit, SNR=5.5) as a function of the peak energy ($E_{peak}$ of the GRB with spectral parameters $\alpha$=-1, $\beta$=-3, located in the center of the field of view of ECLAIRs. ECLAIRs is more sensitive than previous missions to GRB with low $E_{peak}$ (below $\simeq$32 keV, i.e. high redshifted GRB candidates)

CdTe pixels of 1 mm thickness, covering 1024 cm$^2$ of sensitive area. A coded Ta-mask (54 cm side length, composed of 5.4 mm pixels, 30% open fraction, 0.6 mm thickness) is mounted 46 cm above the detection plane and defines, together with the lateral shielding made of Al, Cu and Pb foils, a total field of view of 2 sr on the sky. The CXG detects point sources on the sky in the 4−120 keV energy band, and localizes them with better than 10 arcmin accuracy, based on the technique of detector-plane image-deconvolution using the mask pattern to obtain the sky image.

With its low energy threshold of 4 keV, the ECLAIRs CXG is more sensitive than previous missions to high-redshifted GRB, which are potentially the most interesting ones. This result is based on a detailed Monte-Carlo simulation of the instrument (Geant-4) and the localization algorithm foreseen to be used onboard (Figure 2).

The ECLAIRs on-board Unit for Triggering and Scientific Processing (UTS) uses the CXG data in order to discover the appearance of a gamma-ray source on the sky [4]. It determines the localization of the source to <10 arcmin (see Figure 3), and sends as quickly as possible (within seconds) the trigger information to the satellite to initiate a slew maneuver, as well as to the VHF real-time network from which it is forwarded to the ground observer community. For detailed on-ground analysis, all detected photons are also stored in an on-board mass memory which is dumped to ground via high-bandwidth X-band transceivers after a delay of up to one day.

The second ECLAIRs instrument, the focusing X-ray telescope SXT (soft X-ray telescope), performs further observations of the detected source, in order to obtain a refined localization, which is subsequently transmitted to the VHF real-time network. The SXT starts this GRB follow-up observation once the source has been placed into its 23×23 arcmin$^2$ field-of-view by a satellite slew maneuver ($\simeq$3 min after detection by the UTS). The SXT is composed of 11 nested mirror shells (Au-coated, inherited from the XMM-Newton satellite) with 30 cm of diameter; it has a focal length of 1 m, an energy range from 0.3−2 keV. With its 80 cm$^2$ effective area, the SXT is sensitive to 1 mCrab sources in 100 s, and localizes them to $\simeq$10 arcsec.

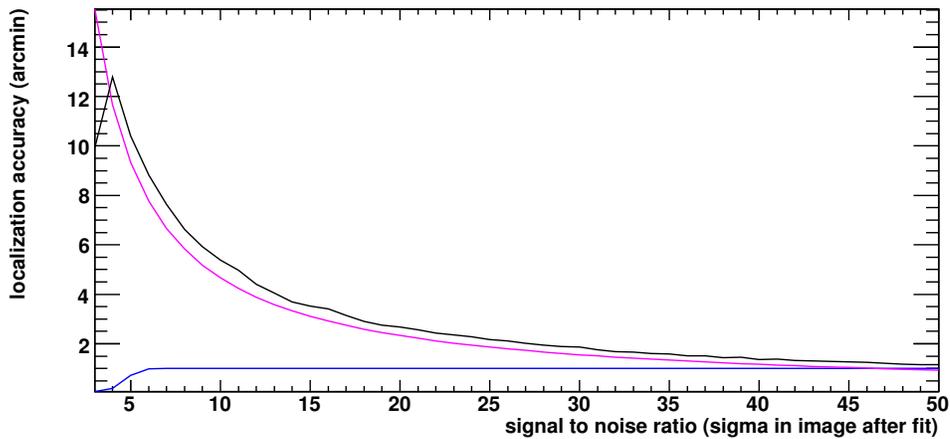

**FIGURE 3.** CXG onboard localization accuracy (arcmin) vs. signal to noise ratio (in sigma) of the reconstructed source, obtained after fitting a 2D Gaussian to the sky image (black; red=theoretical limit) For SNR>5.5, the source is detected in more than 90% of the case (blue values 0 to 1=detection fraction)

The VHF real-time data link to ground is inherited from HETE-2, with a data transfer rate of up to 600 bit/s. It uses a VHF on-board emitter and a network of about 30 VHF ground receiver stations, deployed under the satellite track. This messaging system is used among others to transmit to ground in real-time, within tens of seconds, the localization of GRB detected by ECLAIRs [5].

A dedicated set of ground-based robotic telescopes, operating in the visible and near-infrared bands, will be used to further refine to arcsec accuracy the localization of the GRB (afterglow), and to obtain a photometric redshift estimate.

The overall pointing strategy [6] of the satellite carrying ECLAIRs is to observe the part of the sky roughly opposite to the direction of the sun (for thermal constraints), with avoidance of the Galactic plane and bright X-ray sources such as Sco X-1 (to reduce background). With this scheme, the Earth will be entering the field of view of the instruments every orbit, reducing the overall efficiency by about 30%. However the benefit of this strategy is that almost all detected GRB may have redshift estimates, because they are potentially observable by large spectroscopic telescopes, located above horizon in the Earth tropical zones harboring those telescopes.